\documentclass[conference]{IEEEtran}
\usepackage{hyperref}
\usepackage{tabularx}
\usepackage{supertabular}
\usepackage{capt-of}
\usepackage{xcolor}
\usepackage[sorting=none]{biblatex}
\usepackage{graphicx}

\makeatletter
\renewcommand{\@IEEEsectpunct}{\ \,}
\makeatother

\begin{document}

\title{A Survey on Experimental Performance Evaluation of Data Distribution Service (DDS) Implementations}

\author{\IEEEauthorblockN{Kaleem Peeroo, Peter Popov, Vladimir Stankovic}
\IEEEauthorblockA{Centre for Software Reliability}
\IEEEauthorblockA{City, University of London\\
Kaleem.Peeroo@city.ac.uk, P.T.Popov@city.ac.uk, Vladimir.Stankovic.1@city.ac.uk}}

\maketitle

\begin{abstract}
The Data Distribution Service (DDS) is a widely used communication specification for real-time mission-critical systems that follow the principles of publish-subscribe middleware. DDS has an extensive set of quality of service  (QoS) parameters allowing a thorough customisation of the intended communication. An extensive survey of the performance of the implementations of this communication middleware is lacking. This paper closes the gap by surveying the state of the art in performance of various DDS implementations, and identifying any research gaps that exist within this domain.
\end{abstract}

\section{Introduction}
\label{section:1.introduction}
The Data Distribution Service (DDS) \cite{dds} is a widely used communication specification that has been used for almost two decades. Its usage is embraced by a variety of industries due to its characteristics that facilitate incorporation into real-time, mission-critical, and distributed systems. These industries include, for example, autonomous vehicles, military defence systems, energy management systems, and air traffic management systems. Development of DDS first began in 2001 by Real-Time Innovations (RTI) \cite{rti} and Thales Group \cite{thales_group} before the Object Management Group (OMG) \cite{omg} published the first version of the DDS specification in 2004.

This paper surveys various literature on evaluating the performance of DDS via specific implementations and we are interested in answering the question: How performant is DDS according to various implementations? Since one cannot exactly determine the performance of a specification, whenever this paper refers to the performance evaluation of DDS, it considers the performance of DDS implementation(s). This paper focuses on the performance of DDS compared to other similar communication technologies, the performance of DDS under default QoS settings, and the performance and comparison of various DDS implementations throughout the surveyed literature.

Performance evaluation of DDS is useful for practitioners and researchers as the results demonstrate how different DDS implementations perform under different conditions, and what affects the performance. These results, thus, help understand DDS performance better, and contribute to making decisions that may lead to the usage of DDS over other communication technologies. More specifically, this paper investigates the performance of individual DDS implementations that should benefit the decision about which implementation to choose and under what particular conditions.
Moreover, these implementations change over time and this can therefore affect the ranking in terms of performance. Furthermore, certain implementations may excel in certain scenarios, e.g. implementation X may have better performance when run in virtualised environments than implementation Y, but the opposite might be true for large data sizes in physical or virtualised environment. In this paper we attempt to provide a “clean” comparison of DDS implementations by considering equivalent QoS configurations of these products. 

Section \ref{section:2.data_distribution_service_and_other_communication_technologies} will explain DDS and its components, as well as briefly describe any other communication technologies mentioned throughout the paper. Section \ref{section:3.research_method} will define the approach followed to produce the paper. Section \ref{section:4.results} will discuss the findings of the survey paper. Section \ref{section:5.discussion} will provide an analysis of the survey results while generalising on some themes emanating from the Results section, and finally section \ref{section:6.conclusion} shall conclude on all findings and evaluate whether the survey paper has achieved its objectives.


\section{Data Distribution Service and Other Communication Technologies}
\label{section:2.data_distribution_service_and_other_communication_technologies}
\subsection*{2.1. DDS}
\label{subsection:2.1.dds}
DDS follows the publish-subscribe communication pattern where there exist publishers and subscribers such that publishers publish content that belong to a certain topic and subscribers subscribe to the content of a certain topic. Several publishers and subscribers communicate at any given time, and their respective count can easily be (re-)configured proving effective the scalability of DDS. Publishers use entities known as DataWriters to create the samples to be sent whilst Subscribers use entities known as DataReaders to read these samples sent to the subscriber(s). The most common terms used throughout DDS are \textit{samples}, \textit{instances}, \textit{topics}, and \textit{groups}. A \textit{topic} is a uniquely named data type definition, an \textit{instance} refers to an instance of a topic, a \textit{group} refers to all instances belonging to a DataWriter/DataReader, and a \textit{sample} is an individual item of an instance of a topic which is used interchangeably with message, packet, payload, and datagram.

\begin{figure}
  \includegraphics[width=\linewidth]{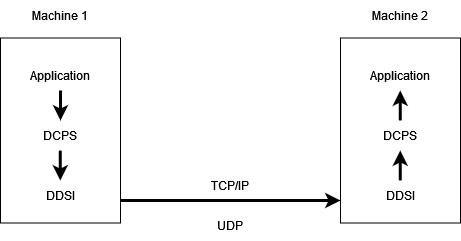}
  \caption{DDS stack traversal.}
  \label{fig:dds_stack_diagram}
\end{figure}

There are two main layers to DDS. The first layer, called the Data Local Reconstruction Layer (DLRL) resides on the same level as the applications and is optional. This layer allows for the ease of integration of DDS with applications. The lower level is called Data-Centric Publish-Subscribe (DCPS) layer which is where all the communication happens. In this layer all the participants, both publishers and subscribers, reside and within them are their constituent DataWriters and DataReaders.
The general process starts within the application where the data is constructed and sent to the DCPS layer where the publishers publish this data. This data is then transferred to the DDS Interoperability wire protocol (DDSI) where the data is then further transported using User Datagram Protocol (UDP)/IP or Transmission Control Protocol (TCP)/IP connections where it is received by another machine and this "stack" is reversed to be used by subscriber(s).

The DDS specification has an extensive range of QoS parameters allowing for numerous configurations specifically customised for various scenarios. These QoS parameters range from data length of the packets being sent throughout the communications, to the history parameter which decides where the participants store packets that have already been sent/received. The DDS specification states that there are 22 parameters of various data types leading to an abundant number of parameter value combinations - too many for one to try out every single combination. If we assume each of the 22 parameters could take a binary value, there is a total of $2^{22}$ combinations which is over four million. In fact, the actual number of possible combinations is likely much greater as many parameters assume more than two possible values.  This is true despite some combinations of parameter values being incompatible, which will decrease the number of possible combinations, and perhaps drastically so. Moreover, there are non-QoS parameters with unlimited values such as data lengths of the messages sent between publishers and subscribers, number of publishers, or number of subscribers.

We have defined the following parameters as parameters of interest due to their impact on performance, and especially the standard measures used such as \textit{latency} and \textit{throughput}: Unicast/Multicast, Data Length, Number of Publishers, Number of Subscribers, Deadline, Durability, Latency Budget, Presentation, and Reliability.

Unicast and Multicast are communication methods and are not considered part of the DDS specification. Unicast represents the communication between two endpoints where there is one sender and one receiver whilst multicast represents the communication from one endpoint to multiple endpoints (but not all endpoints). Theoretically, using the Multicast communication method should increase the throughput as an increased amount of data is being disseminated. By default, DDS does not use Multicast communication though this non-standard QoS parameter can be easily configured, e.g. using a given benchmarking tool.

Data length represents the size of the messages sent between publishers and subscribers and is measured in Bytes. Increasing the data length should theoretically increase both the latency and throughput and various vendor implementations of DDS may use different default values for this parameter.

The number of publishers and the number of subscribers are self-explanatory. Increasing the number of publishers should theoretically increase the throughput depending on the message length and the size of the bandwidth since an increased amount of messages are delivered to the subscribers. Similarly, increasing the number of subscribers should theoretically also result in a related increase.

The Deadline parameter from the DDS standard is used when instances are expected to be periodic (sent/received at certain intervals). The Deadline specifies the amount of time either between writing each sample on a DataWriter or reading each sample on a DataReader. On the publisher side, as soon a sample has been written the timer starts. If the timer reaches the Deadline value stated, then a \verb|DEADLINE_MISSED_STATUS| is sent stating that the next sample could not be written in time. On the subscriber side, as soon as sample has been read the timer starts. If the timer reaches the Deadline value stated, then a \verb|DEADLINE_MISSED_STATUS| message is sent stating that the next sample could not be received in time. By default, the Deadline parameter’s value is set to infinity resulting in a relaxed setting where the packet writing/reading processes are not under any timed constraints.

Durability controls whether readers that join late can access the data that has previously been published. It can take one of four values including \verb|volatile|, \verb|transient_local|, \verb|transient|, and \verb|persistent|. The \verb|volatile| value does not save or deliver formerly published samples. The \verb|transient_local| value saves and delivers published samples if the DataWriter that published the samples still exists. The \verb|transient| value saves and delivers published samples using a memory-based service whilst the \verb|persistent| value saves and delivers old samples using a disk-based service. The impact on the performance of DDS can be affected by the durability since extra packets will either be propagated or not. By default, the Durability QoS parameter assumes the value of \verb|volatile|.

Latency Budget is comparable to the Deadline parameter - it provides a hint to the maximum acceptable delay from the time that the data is written to the time that it is received by the subscribers. This time period indicates the "urgency" of the communication. Since there is no specified mechanism as to how the service should use this hint, different DDS implementations may act differently in terms of performance, and this is one of the aspects that should be explored when evaluating the performance of DDS particularly for desired situations that require this QoS parameter. By default, this parameter has a value of 0 seconds.

The Presentation parameter controls how the data arrives at the DataWriter which includes the sequence of the samples. There are 3 sub-settings: access scope, coherent access, and ordered access. The coherent access setting takes a Boolean value and controls whether the implementation will preserve the groupings of changes made by the publisher indicated by using the operations \verb|BEGIN_COHERENT_CHANGE| and \verb|END_COHERENT_CHANGE|. The access scope setting has 3 values: \verb|instance|, \verb|topic|, and \verb|group|. This setting controls how the samples are ordered, either by topic, instance, or both. If the value is set to \verb|instance|, the queue is ordered for every instance, if the value is set to \verb|topic|, then the queue is ordered per topic, and if the value is set to \verb|group|, then the queue is ordered per topic across all instances belonging to the DataWriter/DataReader within the same respective publisher/subscriber. Finally, the ordered access setting takes a Boolean value and controls whether the implementation will preserve the order of the changes. Theoretically, the latency should be affected if the order is required to remain constant as the implementation may await the arrival of multiple samples following the required order resulting in delays and therefore higher latency.

The Reliability parameter decides whether missed samples are sent again or not. The parameter can take the value of \verb|best effort| or \verb|reliable| (the latter is set by default). \verb|Best effort| does not resend any missed samples and the order of the samples in the history of the DataWriter may not match the history of received samples on the DataReader. On the other hand, if the value of the parameter is set to \verb|reliable|, then any missed samples are resent. If the DataReader does not receive a sample, then the DataWriter blocks the samples ready to be sent to solve the issue. This means that the DataWriter will block all future samples while it attempts to resend the missed sample. The time period elapsed during this blocking can also be configured in order to determine how long the DataWriter can afford to block for. Latency should be affected depending on the value set (\verb|best effort| or \verb|reliable|), and this has been evaluated and discussed further in the paper.

At the time of writing there are several available DDS implementations from various companies. The leading DDS products include Connext DDS from RTI \cite{connnext}, Vortex OpenSplice from ADLINK \cite{opensplice}, FastDDS from eProsima \cite{fastdds}, MilDDS from MilSOFT \cite{mildds}, and OpenDDS from OCI \cite{opendds}. In the surveyed literature RTI product versions ranged from 4.5 to 6, OpenSplice product versions ranged from 5.4.1 to 5.5.1, OpenDDS 3.4.1 was evaluated as well as FastRTPS 1.6.0. Most of these implementations are evaluated within the reviewed papers and their results, and related discussion, will be stated later in this paper, mainly in Section 4 and Section 5 respectively. Despite each implementation following the OMG DDS specification, they differ since the standard does not specify exactly how certain mechanisms must be implemented (e.g. see the description of Latency Budget parameter above).

\subsection*{2.2. Socket-based Communication}
\label{subsection:2.2.socket-based_communication}
Socket-based communication is a two-way communication method for transferring data between two programs on the same network. A socket, which can be identified externally by its socket address being the combination of the transport protocol, IP address, and port number, is bound to an application. The communication process works when the socket on the server process waits for a request from the client. The server first creates an address that clients can use to find the server before waiting for a client's request. The server then carries out the client's request and sends a response back to the client. Sockets are only created during the lifetime of an application process. 

\subsection*{2.3. HLA}
\label{subsection:2.3.hla}
High level architecture (HLA) \cite{hla} is a standard for distributed simulations and is used when constructing larger simulations resulting from the combination of simpler ones. HLA was developed in the 1990s under the leadership of the US Department of Defence and is utilised in distinct domains such as defence and security, and civilian applications.

There are three core components within the HLA system: the framework and rules, the federate interface specification, and the object model template specification. The framework and rules specify 10 architectural rules that all or some federates should follow. Federates are the entities that communicate with the run-time infrastructure and send/receive data from the run-time infrastructure. The federate interface specification specifies services to be given by the run-time infrastructure and these services are given as C++ and Java APIs and Web Services. Finally, the object model template specification specifies the format that HLA object models such as the federation object module should use.

\subsection*{2.4. MQTT}
\label{subsection:2.4.mqtt}
MQ Telemetry Transport (MQTT) \cite{mqtt} is a lightweight publish-subscribe network protocol that usually runs over TCP/IP. It was first authored in 1999 and can be supported by any bi-directional and lossless protocol whilst being designed for networks with very limited bandwidth due to its minimal code. There are two fundamental entities: the broker and the clients. The broker acts as a server and receives all the messages from the clients and re-routes them to other clients. The clients send messages to and receive them from the broker. As in DDS, MQTT uses topics to categorise the data.

\subsection*{2.5. AMQP}
\label{subsection:2.5.amqp}
Advanced Message Queueing Protocol (AMQP) \cite{amqp} is an open standard application layer wire-level protocol for message-oriented middleware and is used in AMQP message brokers such as RabbitMQ to define how the data is communicated between publishers and subscribers. Publisher(s) send data to the broker which stores the messages in a queue before these messages are pulled by the subscriber(s). AMQP was created in 2003 by John O'Hara at JPMorgan Chase in London and AMQP 1.0 was released by the AMQP working group on the 30 October 2011.

\subsection*{2.6. ZeroMQ}
\label{subsection:2.6.zeromq}
ZeroMQ, or ZMQ \cite{zmq}, is an asynchronous messaging library used in distributed and concurrent applications. It provides a message queue and can run without a broker. The ZMQ API provides sockets where each one represents a many-to-many connection between endpoints. One of four messaging patterns is required to be used and these patterns include: request-reply, publish-subscribe, push-pull (pipeline), and exclusive pair. The request-reply pattern connects a set of clients to a set of services, the publish-subscribe pattern connects publishers to subscribers, the push-pull (pipeline) pattern connects nodes in a fan-out/fan-in pattern, and the exclusive pair pattern connects two sockets in an exclusive pair.

\section{Research Method}
\label{section:3.research_method}
The overall objective of this survey is to identify what has been researched in the field of experimental evaluation of DDS performance. This includes performance of DDS under default QoS conditions, when compared to other communication technologies, and under specific use cases such as air traffic management, video streaming, etc. To achieve this, we have applied a systematic literature review which is a well-defined and precise method of identifying, evaluating and analysing studies related to the subject of interest. A systematic literature review here aids in identifying any research gaps to create a foundation whereby further investigative research work can be realised. The research method, from which this survey paper stems from, is explained in \cite{systematic_literature_reviews_in_software_engineering_a_systematic_literature_review_2009}. This method has been used throughout other survey papers related to DDS such as \cite{obstacles_in_data_distribution_service_middleware_a_systematic_review_2017}. There exist six main points of the proposed approach, which have been tailored slightly to suit our survey:
\begin{enumerate}
    \item Identification of the research question
    \item Identification of a search strategy
    \item Identification of the study selection criteria
    \item Data extraction from studies
    \item Data summary from data extractions
    \item Critical analysis and theme identifications
\end{enumerate}
The structured process commences with the identification of the research question and any sub-questions. This is in preparation for the following step: identifying the scope, since the questions aid in the creation of the scope for the papers of interest. Since the focus of this survey paper is the experimental performance assessment of DDS, the scope is limited to just that and any papers regarding other (related) topics, for example security vulnerabilities of DDS, have been excluded. Once the scope is identified, we can then create a search strategy to collect all (to the best of our knowledge) the papers of interest. This strategy defines where searches will be made and what the search terms are to reproduce these search results if required. After the completion of this step, we are left with a collection of studies from various literature, e.g. journals, conferences and symposiums proceedings, etc. From this initial collection we filter out any unrelated papers where these filtering rules are identified later in section \nameref{subsection:3.4.selection_criteria}. This produces a collection from which we extract the data that benefits in answering the given research question. Once the data is extracted and summarised, we critically analyse the literature and identify themes where we group the papers according to their theme. We complete the survey paper with identification of research gaps and possible future work.

\subsection*{3.1. Research Question}
\label{subsection:3.1.research_question}
The process begins with the identification of the research question. This paper focuses on surveying results about experimental performance evaluation of DDS. Since DDS itself is a specification, one cannot meaningfully evaluate its performance and we can instead use a sample of the most reputable DDS implementations to do so. This has led to the research question: "How performant is DDS according to its implementations?" which defines a scope for the rest of this survey involving two main sub questions. 

A research sub-question is as follows: "What is the difference in performance of various DDS implementations?". Various DDS implementations have been evaluated in the surveyed literature to determine which implementations excel under which conditions. As mentioned previously within the introduction to section \nameref{section:2.data_distribution_service_and_other_communication_technologies}, a complication is that various DDS implementations have different versions each of which is likely to exhibit a different performance output. We rely on the reviewed literature to provide results for various versions of a given DDS implementation.

The following research sub-question "How performant is DDS compared to other communication standards?" focuses on the performance comparison of DDS to other communication standards mentioned in section \nameref{section:2.data_distribution_service_and_other_communication_technologies}. This question allows us to investigate the performance of DDS in comparison with other communication technologies and the authors of the related papers indeed used implementations of the various communication standards to carry out the performance comparison.

From these research sub-questions we generalise on the results to determine how performant the DDS specification is according to the selected implementations. We also determine how these implementations perform compared to the surveyed implementations of other communication standards

\subsection*{3.2. Scope Identification}
\label{subsection:3.2.scope_identification}
Once the research question has been identified, a scope had to be clarified to define what type of papers would be pertinent for this review. The main subject of this paper is performance evaluation of DDS implementations, assessed through various applications and under different settings. We have collected possibly all (to the best of our knowledge and efforts) papers between 2003 and 2021 that are related to DDS and its performance evaluation. The year 2003 has been particularly chosen as this was when the OMG released the DDS specification and 2021 is the time of which this paper is being written. The scope will also contain all papers that contain performance metrics related to DDS subjects including the application of DDS with other technologies as well as evaluation of the performance of DDS within virtualised environments.

\subsection*{3.3. Search Strategy}
\label{subsection:3.3.search_strategy}
The search strategy includes any means of identifying potential papers of interest. The main search strategy used included several search terms applied to Google Scholar \cite{google_scholar}. We also applied these search terms to IEEE Xplore, ACM Digital Library, Wiley, Science Direct and Springer including:

\begin{itemize}
    \item "data distribution service performance evaluation"
    \item "data distribution service performance"
    \item "dds performance"
    \item "data distribution service performance assessment"
    \item "dds performance assessment"
    \item "opensplice dds"
    \item "rti dds"
\end{itemize}

A total of 46 pieces of literature were collected including survey papers and PhD theses. The relatively low number of the literature collected points to a lack of research in this area and this number decreases in the next step once the selection criteria are applied.
    
\subsection*{3.4. Selection Criteria}
\label{subsection:3.4.selection_criteria}
The selection criteria filter out the literature of no interest. We used these criteria to further focus on papers that aid in answering the research question and to remove any papers that are not pertinent. The criteria included accepting papers that are written in English (some papers were written in Korean whilst their abstracts being written in English), accepting papers that included performance test results in the form of graphs or tables, accepting papers that did not change how the DDS specification worked to gain performance improvements (with some exceptions since RTI has implemented features such as AutoThrottle which is an experimental feature that allows the user to configure a Data Writer to automatically adjust its writing rate and send window size to provide the best latency/throughput trade-off while system conditions change as mentioned in section 10.4 of \cite{rti_connext_dds_core_libraries_users_manual}).

After applying the filtering to the 46 papers, the list narrowed down to 29. Out of these, 17 papers were published in conference proceedings,  10 in journals and 2 had unidentifiable sources (originally found as PDF files resulting from a Google Scholar search).

\subsection*{3.5. Data Extraction}
\label{subsection:3.5.data_extraction}
After collecting a considerable number of papers, we extracted the useful information related to the research question and we had to identify the common data that occurred frequently throughout all papers. This would include items of interest to the performance evaluation of DDS implementations as listed below:
\begin{itemize}
    \item Data Lengths
    \item Parameters Investigated During the Test (including both QoS and non-QoS parameters)
    \item Number of Publishers
    \item Number of Subscribers
    \item Products Evaluated
    \begin{itemize}
        \item Product Version
    \end{itemize}
    \item Publication
    \item Publication Type
    \item Paper Subject
    \item Number of Experimented Parameters
    \item Number of Products Evaluated
\end{itemize}
For each paper, we identified exactly what performance measurements were taken and then recorded the conditions of the tests under which the measurements were obtained. This was followed by the summary of the tests along with its corresponding conditions such as QoS values and various items associated with its description.

The data lengths recorded are the sizes of the packets sent and received during the tests whilst the Number of Publishers and Subscribers are self-explanatory: they are the respective participant amounts present in a given test.

The parameters investigated during the tests include both QoS and non-QoS-related parameters where the parameters’ values were explored and experimented with. For example, if two tests were run, one with Best Effort and the other with Reliable setting, then the parameter of interest would be Reliability. Table \ref{parameter_table} lists the parameters gathered with their corresponding descriptions.

\begin{table}[h]
    \begin{center}
        \begin{tabular}{|p{3cm}|p{5cm}|}
            \hline
            \textbf{Parameter} & \textbf{Description} \\
            \hline
            AutoThrottle & Automatically adjusts the writing rate and the send window size. \\
            \hline
            Batching & Batches samples together to send instead of sending one sample at a time. \\
            \hline
            Deadline & Amount of time either between writing each sample on a DataWriter or reading each sample on a DataReader.\\
            \hline
            Destination Order & Controls what sample is chosen when multiple DataWriters send the same sample to a DataReader.\\
            \hline
            Durability & Controls whether data reader that join late can access already published data. \\
            \hline
            History & Controls how many samples to be stored on either the DataWriter or DataReader. \\
            \hline
            Latency Budget & Provides a hint to the maximum acceptable delay from the time the data is written to the time it is received by the subscribing applications. \\
            \hline
            Life Span & Controls how long a sample can be stored in cache for being deleted. \\
            \hline
            Liveliness & Controls whether a DataWriter is "alive" in order to be used by the ownership parameter. \\
            \hline
            Maximum Transmission Unit (MTU) & Size of the largest packet size that can be sent on the network. \\
            \hline
            Message Size & Data length, size of the sample.\\
            \hline
            Multicast & Controls whether the data can be sent one at a time or to multiple nodes at once.\\
            \hline
            Number of Nodes & Amount of nodes on the network containing publishers and subscribers.\\
            \hline
            Presentation & Controls how the data arrives at the data reader.\\
            \hline
            Publisher Amount & Number of publishers used in the test.\\
            \hline
            Publishing Rate & Samples published per second.\\
            \hline
            Reliability & Controls whether missed samples are resent or not. \\
            \hline
            Subscriber Amount & Number of subscribers used in the test.\\
            \hline
            Transport Priority & Optional QoS parameter that describes the priority of the sample being sent.\\
            \hline
            TurboMode & Adjusts the number of bytes in a batch at runtime according to current system conditions.\\
            \hline
        \end{tabular}
    \end{center}
    \caption{Parameters and their descriptions.}
    \label{parameter_table}
\end{table}

DDS implementations that were evaluated based on the surveyed papers are as follows: Vortex OpenSplice, RTI's DDS, FastDDS, and OpenDDS. The DDS products allow users to create required DDS applications and should not be confused with associated benchmarking tools, such as PerfTest \cite{rti_perftest} provided by RTI to evaluate and monitor DDS performance (of their, and other, products), via assessing  latency, throughput, CPU utilisation, etc.

\subsection*{3.6. Data Summary}
\label{subsection:3.6.data_summary}
Categorisation of the papers began by classifying the general themes of the papers followed by documentation of the category. This led to some categories containing only a small number of, or just one single, piece of literature indicating that these were too specific. Therefore, once all the papers were surveyed and all categories were identified, we grouped the papers in more meaningful, generic categories.
The initial list of categories included the following:

\begin{itemize}
    \item Security Evaluation
    \item DDS Translations
    \item Performance of DDS
    \item Robustness of DDS
    \item Applications of DDS
    \item Obstacles of DDS
    \item Modelling of DDS
    \item Performance of Applied DDS
    \item Performance of Virtualised DDS
\end{itemize}

Once all the papers were surveyed, we decided to focus on the following 3 main categories due to their relation to the evaluation of the performance of DDS:

\begin{itemize}
    \item Performance of DDS
    \item Performance of Applied DDS
    \item Performance of Virtualised DDS
\end{itemize}

\textit{Performance of DDS}
\label{textit:performance_of_dds}
\par This is the main category of interest for this survey paper. These papers evaluated the performance of DDS, including comparison of DDS performance against other publish-subscribe technologies, compared multiple DDS implementations and experimented how specific different QoS and non-QoS parameters affected the performance of DDS.
\linebreak

\textit{Performance of Applied DDS}
\label{textit:performance_of_applied_dds}
\par Authors who applied DDS to certain scenarios or integrated it with other technologies will have their papers placed under this category. This differs from the previous type as with the previous types various test conditions were chosen depending on what values were being investigated and how these values affected the performance. On the other hand, this category specifically looked at test conditions that were customised for certain scenarios whereby the test parameters were carefully considered to produce an optimal performance result.
\linebreak

\textit{Performance of Virtualised DDS}\\
\label{textit:performance_of_virtualised_dds}
\par This category contains all papers that evaluated the performance of DDS when run in virtualised environments which contains scenarios where DDS is tested on virtual machines.
\linebreak

\subsection*{3.7. Critical Analysis and Theme Identification}
\label{subsection:3.7.critical_analysis_and_theme_identification}
At this stage of the survey, we had collected a list of papers, before extracting and summarising the relevant data. From this point we had to analyse the results and discover any themes which resulted in the following list:

\begin{itemize}
    \item Comparison of DDS with other Publish-Subscribe Technologies
    \item Evaluation of Multiple DDS Implementations
    \item Evaluation of a Single DDS Implementation
    \item Evaluation of Unnamed DDS Implementations
    \item Evaluation of DDS Implementations in a Virtualised Environment
\end{itemize}

The papers belonging to each theme were further explained in the section \nameref{subsection:4.2.categories_investigated} including comparison of DDS with other publish-subscribe technologies, evaluation of multiple DDS implementations, evaluation of a single DDS implementation, evaluation of unnamed DDS implementations, and evaluation of DDS implementations in a virtualised environment. Some papers belong to more than one theme as they contained multiple types of tests. In these cases, we have extracted the parts that correspond to a given theme and included it in the related section. Thus a paper’s results can be found in more than one subsection as seen in section  \nameref{subsection:4.2.categories_investigated}.

Throughout the extraction of the test results, we have focused on three main performance metrics: \textit{latency}, \textit{throughput}, and \textit{jitter}. This has mainly stemmed from the fact that these were the main metrics provided by majority of the papers. This was to be expected since these 3 metrics are vital in evaluating the performance of any publish-subscribe technology. The latency is usually measured in milliseconds though sometimes when values are small enough, the microseconds units are used. It should also be stated that some papers measured the latency using the system clocks where both publisher(s) and subscriber(s) resided, and used certain techniques to increase synchronicity between the timings. On the other hand, other papers measured the latency on the publisher side only, by taking first the measurement a data sample is published by a given publisher, and subsequently when the acknowledgement that a subscriber has obtained the sample is received by the publisher. In this way the two timestamps are taken on the same system where the publisher resides. This value, known as the round-trip time (RTT) would then be halved resulting in a one-way latency. In either way, the true one-way latency is not really obtained since the value cannot exactly be attained due to lack of true synchronicity or the fact that the second "half" of the two-way measurement could affect the value of the actual one-way latency and this therefore means that a "fair" comparison cannot exactly be made which reinforces the gap in the research about the accuracy of the latency measurements.However, there are well known approaches for attempting to get as close as possible to this true synchronicity such as \cite{time_clocks_and_the_ordering_of_events_in_a_distributed_system_1978}.

\hfill\\
\textit{Comparison of DDS with Other Publish-Subscribe Technologies}\\
\label{textit:comparison_of_dds_with_other_publish-subscribe_technologies}
A total of 6 papers were classified under this theme and processing the information from the papers under this theme was straightforward. We identified the different tests within a single paper for the corresponding publish-subscribe technologies and summarised the test results for each test. With these summarised findings we generalised and identified common results and grouped them together before recording it within this paper.

\hfill\\
\textit{Evaluation of Multiple DDS Implementations}\\
\label{textit:evaluation_of_multiple_dds_implementations}
A total of 7 papers were classified under this theme and processing the literature from this section took more effort than the previous theme. For each paper we identified all the tests carried out, then for each test we identified the parameter values and the implementation used. We then grouped the parameter values per test per paper and identified common parameters per implementations before grouping the parameter results per implementation leading to the identification of further subheadings for section 4.2.2 of the results section.

\hfill\\
\textit{Evaluation of a Single DDS Implementation}\\
\label{textit:evaluation_of_a_single_dds_implementation}
A total of 11 papers were classified under this theme. Following similar processes as the previous theme, for each paper we identified the tests carried out. For each test we gathered the main parameter of interest. This was the parameter that really affected the outcome of the test results and would be used as the subheadings for section 4.2.3 of the results section. As well as this we collected the parameters and their values used in the tests. We finally gathered the results in terms of the latency, throughput, and jitter as well as the DDS implementation used. Once we gathered all this data in a table format, we then re-organised the papers according to parameter value (see sub-section \nameref{subsubsection:4.2.3.evaluation_of_a_single_dds_implementation}) .

\hfill\\
\textit{Evaluation of Unnamed DDS Implementations}\\
\label{textit:evaluation_of_unnamed_dds_implementations}
A total of 5 papers were classified under this theme. To process this literature, we followed the same process (with some minor streamlining such as not recording the products evaluated and their versions due to the lack of this information) as the previous theme. For each paper we gathered the tests, their configurations, and their results according to latency, throughput, and jitter. We then categorised these tests according to their purpose and these categories included items such as scalability, simulation, and unicast vs multicast. From there it was easy to summarise the results and findings into the results section since there were only three categories of tests.

\hfill\\
\textit{Evaluation of DDS Implementations in a Virtualised Environment}\\
\label{textit:evaluation_of_dds_implementations_in_a_virtualised_environment}
A total of 6 papers were classified under this theme and for the processing of this literature, we copied the process of the previous theme with a minor addition. We collected "general results" on top of the latency, throughput, and jitter results as the papers gave its findings in terms of items such as invocations per second, virtualisation overhead, and blocking time for small messages.
\section{Results}
\label{section:4.results}
In this section we have summarised all the results from the surveyed literature and have categorised the results accordingly into five assorted categories as well as have mentioned all the results of the tests executed throughout the literature. Each of these categories have also been further classified according to the pertinent themes that were extracted. It should be noted that when a paper mentions that the enclosed tests are characterised with the "default" QoS settings, it is referring to the default values of the QoS parameters stated by the OMG DDS specification. 

\subsection*{4.1. Overview of Selected Studies}
\label{subsection:4.1.overview_of_selected_studies}
A total of 29 papers were thoroughly reviewed throughout this section. From all these papers, the number of publishers involved within the reviewed experiments ranged from 1 to 45 whilst the subscriber amount ranged from 1 to 100. Furthermore, 6 papers compared the performance of DDS with other publish/subscribe technologies such as Message Queueing Telemetry Transport (MQTT), 7 papers compared the performance of numerous versions of various DDS implementations, 11 papers evaluated the performance of specific DDS implementations where the tests focused on certain experimental conditions, 4 papers evaluated the performance of DDS implementations without mentioning their names or versions, and finally, 6 papers evaluated the performance of DDS when executed in a virtualised environment though 3 of these papers had similar authors and content. Its noticeable that the sum of the mentioned papers comes to 34 which is greater than 29 and reinforces the point that multiple reviewed papers and their corresponding test summaries may occur in more than one theme. The test with the most widely ranged configuration encompassed 8 different test variables (including QoS parameters) and the most popular test variable was the message size which was evaluated in 13 of the papers followed by the reliability parameter being evaluated in 8 papers. RTI DDS was the most popular vendor evaluated with 19 papers investigating its performance whilst FastRTPS was the least popular product with only 4 papers investigating its performance ignoring any DDS implementations that were not mentioned in the reviewed literature. There was a range of DDS implementations evaluated including RTI DDS, FastRTPS, OpenDDS, and OpenSplice.

\subsection*{4.2. Categories Investigated}
\label{subsection:4.2.categories_investigated}
Overall, the papers have been put into five categories which are distinct from what was mentioned in \nameref{subsection:3.6.data_summary} as these five categories dive into the mentioned categories of \nameref{subsection:3.6.data_summary} further and additionally are more unequivocal:

\begin{itemize}
    \item Comparison of DDS with other Publish-Subscribe Technologies
    \item Evaluation of Multiple DDS Implementations
    \item Evaluation of a Single DDS Implementation
    \item Evaluation of Unnamed DDS Implementations
    \item Evaluation of DDS Implementations in a Virtualised Environment
\end{itemize}

\hfill
\subsubsection*{4.2.1. Comparison of DDS with other Publish-Subscribe Technologies}
\label{subsection:4.2.1.comparison_of_dds_with_other_publish-subscribe_technologies}
\hfill\\
This section will identify how DDS compares with other publish-subscribe technologies in terms of performance. There are numerous other publish-subscribe technologies that compete with DDS and this section intends to clarify which technologies are more performant and using what test parameter configurations. The listed non-DDS technologies within this section were specifically chosen according to the reviewed literature as these were the explicit technologies mentioned.

\textit{4.2.1.1. Traditional Socket-based Communication}\\
\label{textit:4.2.1.1.traditional_socket-based_communication}
Traditional socket-based communication is one of the most popular network communication methods used throughout the world and significantly outperforms DDS in all conditions tested. An evaluation and comparison of the performance of DDS with the traditional socket-based communication was performed by \cite{data_distribution_service_for_industrial_automation_2012}. When varying the message sizes the test results showed that DDS was more sensitive to bigger messages and was significantly slower than socket-based communication which had a latency ranging from 62.5 to 112.5 microseconds while DDS had a latency ranging from 50 to 400 microseconds resulting in TCP having less variation in its latency. Further work could be carried out to investigate under what exact situation DDS outperforms socket-based communication given the minimum values of the DDS latency range.

\textit{4.2.1.2. HLA}\\
\label{textit:4.2.1.2.hla}
High level architecture (HLA) is a standard for distributed simulation and its performance was evaluated and compared to DDS in \cite{addressing_the_challenge_of_distributed_interactive_simulation_with_dds_2010} where the results show that between 0 to 3000 bytes, the two implementations of the standards had similar latency, though after 3000 bytes, DDS had slightly lower latency. Furthermore, the results state that DDS outperforms HLA in terms of jitter as well as the fact that DDS has a higher throughput as shown in Table 2 of the paper. We see a huge difference between the two communication technologies. Little is known as to why DDS experiences a huge rise in throughput for 5 kilobyte messages and we speculate that this could this be because of the best effort quality of service setting though further exploration and investigation is required. The messages aren't bigger than the User Datagram Protocol (UDP) packet limit. In fact, roughly 12 messages can fit into 1 UDP datagram. If we look at the message size it jumps from 1,000 to 5,000 bytes, a multiple of 5. But looking at the throughput of DDS, it jumps from 112 to 800 megabytes per seconds, almost a multiple of 8. Further investigation is needed to explain this behaviour.

\textit{4.2.1.3. MQTT}\\
\label{textit:4.2.1.3.mqtt}
MQTT is a lightweight publish-subscribe technology used in a wide variety of industries such as automotive, manufacturing, and telecommunications and multiple papers have compared the performance of MQTT with DDS. MQTT has been mentioned in \cite{industrial_messaging_middleware_standards_and_performance_evaluation_2020} as not being reliable due to its architecture where there exists a single point of failure. Three types of tests were run in \cite{a_study_of_publish_subscribe_middleware_under_different_iot_traffic_conditions} for the mentioned technologies: high frequency, periodic, and sporadic. In the high frequency test (where publishers sent unlimited data to subscribers using default QoS settings), the results show that when the messages where smaller than 1 kilobyte, MQTT had greater throughput than DDS. In fact, DDS had the lowest throughput out of ZeroMQ and MQTT. The authors that attained these results explain that this outcome was achieved because ZeroMQ is built on top of the socket layer which is lower than DDS and this culminates into less time executing application-level data processing. However, when the messages were larger than 1 kilobyte, DDS had greater throughput than MQTT and had the largest throughput of all the tested technologies within this paper. These were the results with 1 publisher and 1 subscriber. In the case where 1 publisher disseminated data to several subscribers, the results indicate that whilst DDS outperforms MQTT in terms of throughput, ZeroMQ drastically outperforms both DDS and MQTT when messages were under 1 kilobyte. Messages greater than 1 kilobyte were still dominated by ZeroMQ but with a smaller gap in competition with the performance of DDS. The comparison of these results can be seen in Figure 1 of the mentioned paper. The high frequency test results conclude that DDS outperforms MQTT when 1 publisher publishes to 1 or 7 subscribers with default QoS settings.

In the periodic tests, where 1 publisher exchanged messages with 1 subscriber, the authors varied the data size (taking values of 64 bytes, 2 kilobytes, and 32 kilobytes) as well as the publishing rate (taking values of 200, 400, 600, 800, and 1000 samples per second). The results show that for 64-byte messages, the average of latency of DDS was always lower than that of MQTT because the MQTT broker needs more time to process the traffic as the rate is increased. When the broker cannot keep up with the publisher and the write buffer of the publisher is exhausted, the writing from the publisher is blocked. In all the publishing rates except for 400 samples per second, the average latency of ZeroMQ was lower than that of DDS where these trends were also witnessed for 2 kilobyte messages. For 32 kilobyte messages, the results drastically changed. For publishing rates greater than 200 samples per second, the average latency of DDS was greater than MQTT's average latency and ZeroMQ had the largest average latency of the 3 technologies. In the case where the publishing rate was 200 samples per second, ZeroMQ had the lowest average latency closely followed by DDS, and MQTT had the largest average latency of around 8ms. In summary, the results tell us that DDS outperforms MQTT in terms of latency for messages of 64 bytes and 2 kilobytes in length while the case is the opposite for messages that are 32 kilobytes in length and the comparison of these results are graphically represented in Figure 2 of the mentioned paper.
In the sporadic tests, where the author simulated the scenario in which part of the system needs to suddenly send a large amount of data in one go, the results show us that for all message sizes experimented on, DDS outperformed MQTT significantly as MQTT is more sensitive to bursts of data streams because of the broker-centric architecture. Further confirmation that DDS outperforms MQTT is shown in \cite{opc_ua_versus_ros_dds_and_mqtt_performance_evaluation_of_industry_4.0._protocols_2019} where comparison of Figure 1c and 1g of the mentioned paper shows that the performance gap is not too significant. This trend is also confirmed in \cite{performance_evaluation_of_iot_protocols_under_a_constrained_wireless_access_network_2016} where the paper mentions that even though DDS consumes twice the bandwidth of MQTT (since DDS must generate at least twice the number of control packets as MQTT) it still outperforms MQTT for latency under all conditions. The paper even shows that in a test looking at experienced packet loss, DDS loses no packets while MQTT does. In low quality wireless network tests, DDS outperforms MQTT for latency, packet loss, and narrow network bandwidth cap.

\textit{4.2.1.4. Other Publish-Subscribe Technologies}\\
\label{textit:4.2.1.4.other_publish-subscribe_technologies}
Other communication technologies evaluated include RabbitMQ and AMQP which were found to be the most stable. However, they were not the fastest as seen in \cite{industrial_messaging_middleware_standards_and_performance_evaluation_2020}. ZeroMQ is another technology that consistently outperforms DDS for message sizes under 1 kilobyte as well as when 1 publisher and 7 subscribers are used according to \cite{a_study_of_publish_subscribe_middleware_under_different_iot_traffic_conditions} further described in \nameref{textit:4.2.1.3.mqtt}. Furthermore, OPC UA's open62541 outperforms DDS whilst OPC UA's client/server performs worse than DDS as seen in \cite{opc_ua_versus_ros_dds_and_mqtt_performance_evaluation_of_industry_4.0._protocols_2019} which concludes that open26541 of OPC UA and FastRTPS of DDS delivers high performance being the two highest performing technologies. 

\hfill
\subsubsection*{4.2.2. Evaluation of Multiple DDS Implementations}
\label{subsubsection:4.2.2.evaluation_of_multiple_dds_implementations}
\hfill\\
The purpose of this section is to extract, analyse and compare the findings of the numerous implementations of DDS that have been evaluated and compared throughout the retrieved collection of relevant scientific literature. A total of six pieces of literature have been included within this section and all these papers have evaluated at least two different implementations of DDS. Since these papers have been published at different times the names of these DDS implementations may have changed and therefore, to ensure consistency throughout the results, we refer to the ADLINK implementation as OpenSplice, the OCI implementation as OpenDDS and finally the RTI implementation as RTI DDS (though in many cases it was referred to by its most recent title: Connext).

\textit{4.2.2.1. Data Length}\\
\label{textit:4.2.2.1.data_length}
Overall, all 6 papers evaluated the performance of DDS when varying the values of data length ranging from 2 bytes to 4 megabytes. Typical UDP packets are 64 kilobytes in size and therefore any messages bigger than this are split into multiple packets and observing how this affects the performance would be interesting since theoretically, large data lengths should increase the latency and throughput.

The piece of literature \cite{data_distribution_services_performance_evaluation_framework_2018} tested data lengths ranging from 100 bytes to 100 kilobytes and found that OpenSplice was slower than OpenDDS in terms of latency and jitter (lower latency and higher jitter). This paper also compares the performance of OpenSplice, OpenDDS and FastRTPS with data lengths ranging from 100 bytes to 1 megabyte where the results from this comparison show us that OpenSplice has the highest latency, followed closely by FastRTPS's latency, and OpenDDS had the lowest latency of the three implementations whilst latency increases linearly as the time progresses for all the products. In summary, this result shows us that whilst the latency linearly increases for all implementations (as the data length increases), OpenSplice has the highest latency, followed by FastRTPS and OpenDDS with the lowest latency. It must be noted that the paper mentions that all the implementations were tested in "safe" mode, meaning that all messages were stored in memory the entire time as well as the fact that every message was followed with a confirmation about its delivery and this may be the reason why the processing time of each next message went up.

Performance evaluation of OpenSplice and RTI DDS was carried out in \cite{dds_a_performance_comparison_of_opensplice_and_rti_implementations_2013} whilst varying the data length between 128 bytes and 512 kilobytes. The results in figure 4b of the mentioned paper shows that OpenSplice has a higher throughput up to the 4-kilobyte mark. From that point onwards RTI DDS significantly outperforms OpenSplice in terms of throughput since RTI DDS has a self-contained decentralised architecture. The two implementations had almost the same sample rate though the RTPS protocol implementation can get better bandwidth. OpenSplice crashes once a data length of 112 kilobytes is experimented on while RTI continues performing and the authors explain that this could be because of the saturation of the DCPS daemon continuing to consume RAM and CPU until the crash due to such a heavy system overload. It is also stated that after 64 kilobytes, the message must be split into multiple UDP packets which is the reason why the throughput remains the same while the rate decreases. Data is being transported and DDS is waiting for data to be sent before it starts writing again and RTI DDS's defragmentation seems to be highly optimised as mentioned by the authors of the paper. The paper also mentions that RTI DDS saturates the bandwidth when a data length of 16 kilobytes was used. In summary, these results show that OpenSplice outperforms RTI DDS up to data lengths of 4 kilobytes and for data lengths greater, RTI DDS outperforms OpenSplice in terms of throughput.

Performance of OpenSplice for data lengths ranging from 256 bytes to 4 megabytes amongst many other QoS configurations was carried out in \cite{exploring_the_performance_of_ros2_2016}.  Figure 7 of the mentioned paper shows that the latency of OpenSplice is constant for data lengths up to 4 kilobytes where it begins increasing non-linearly to the point where it reaches around 6 milliseconds for data lengths of 64 kilobytes. Meanwhile, figure 9 of the mentioned paper surprisingly shows a different outcome: despite the increase in data length, the latency remains constant. This is because DDS requires marshalling various configurations and decisions for the QoS policy. The authors mention that they observe a trade-off between latencies and quality of service policy regardless of the data size as well as the fact that the latencies are predictable and small even though the QoS configurations produced overhead. Figure 9 shows us that the data length does not affect OpenSplice's latency up to 64 kilobytes whilst figure 10 of the mentioned paper shows a non-linear increase up to 4 megabytes before a sharp increase in the latency when large data is split into several datagrams. From these results we can summarise that OpenSplice's latency increases while the data length increases as with all DDS implementations and the degree of the increase is determined by whether the nodes are local or remote. Specifically, OpenSplice seems to handle small data lengths significantly better than larger data lengths.

Evaluation and comparison of the performance of OpenSplice and RTI DDS  was carried out in \cite{security_and_performance_trade_offs_for_data_distribution_service_in_flying_ad_hoc_networks_2019} for data lengths ranging from 100 kilobytes to 1000 kilobytes and the results shown in figure 6 of the mentioned paper show us that OpenSplice's latency is significantly greater than its competitors RTI DDS and OpenDDS which both have similar latency values. From these results we can deduce that OpenSplice is not well-suited for large data lengths, a solidifying trend.

It has been clearly shown that for larger data lengths, OpenSplice is outperformed by its competitors. However, the question in focus now is how does it perform in comparison for smaller data lengths and are the differences significant? We know from \cite{dds_a_performance_comparison_of_opensplice_and_rti_implementations_2013} and \cite{exploring_the_performance_of_ros2_2016} that OpenSplice's performance is underwhelming for data lengths greater than 4 kilobytes, so let us investigate the comparison of OpenSplice's performance for data length below 4 kilobytes.

OpenSplice outperforms RTI DDS in \cite{dds_a_performance_comparison_of_opensplice_and_rti_implementations_2013} in terms of throughput for data lengths between 128 bytes and 512 kilobytes mainly because of batching. Figure 9 of \cite{exploring_the_performance_of_ros2_2016} shows us that OpenSplice outperforms RTI DDS with a lower latency for all data lengths from 256 bytes to 4 kilobytes though it should be noted that OpenSplice is using the “reliable” reliability setting while RTI DDS is using the “best effort” reliability setting in this graph which is an example of an "unfair" comparison. Therefore, from these results we can summarise that OpenSplice outperforms RTI DDS in terms of throughput for data lengths smaller than 4 kilobytes.

At this point, OpenSplice outperforms RTI DDS for smaller data lengths and is outperformed by all implementations for larger data lengths. In the following paragraphs we will investigate how the other implementations compare with each other when the data lengths vary.

Figure 4b of \cite{data_distribution_services_performance_evaluation_framework_2018} shows us that for data lengths ranging from 100 bytes to 10 kilobytes, OpenDDS outperforms FastRTPS with a lower latency. Meanwhile, for data lengths ranging from 10 kilobytes to 1 megabyte, the opposite is the case; FastRTPS outperforms OpenDDS with a lower latency. Therefore, it seems that OpenDDS is also better suited for smaller data lengths up to 10 kilobytes whilst FastRTPS is better suited for larger data lengths ranging from 10 kilobytes to 1 megabyte.

Figure 6 of \cite{security_and_performance_trade_offs_for_data_distribution_service_in_flying_ad_hoc_networks_2019} shows us that between 100 to 500 kilobytes and from 700 to 1000 kilobytes, RTI DDS subtly outperforms OpenDDS with slightly lower delays. A strange result is shown between 500 and 700 kilobytes since OpenDDS outperforms RTI DDS. Comparison of the performance of RTI DDS and OpenDDS was done in \cite{performance_assessment_of_omg_compliant_data_distribution_middleware_2008} and shows us that between 32 bytes and 8 kilobytes, RTI DDS has a lower latency though the difference is not significant, and the jitter tells us that RTI DDS has more stable behaviour than OpenDDS. For data lengths greater than 8 kilobytes, OpenDDS outperforms RTI DDS where the difference in performance increases with the data length. Meanwhile, the authors in \cite{opc_ua_versus_ros_dds_and_mqtt_performance_evaluation_of_industry_4.0._protocols_2019} evaluated OpenDDS and FastRTPS for data lengths ranging from 2 bytes to 32 kilobytes and mentioned that OpenDDS is considerably slower than FastRTPS throughout the tests.

In conclusion, the literature has shown that for smaller data lengths ranging from 1 byte to 4 kilobytes, OpenSplice is very performant as well as OpenDDS (though up to 10 kilobytes in this case). The performance of these implementations substantially worsens as the data lengths increase. For larger data lengths ranging from 10 kilobytes onwards, RTI DDS and FastRTPS are more performant and better suited.

\textit{4.2.2.2. Reliability}\\
\label{texit:4.2.2.2.reliability}
Overall, a total of 3 papers evaluated the performance of DDS when best effort and reliable settings of the reliability parameter were used though we focus on two of these papers due to one of the papers focusing on other publish-subscribe technologies. Theoretically, the reliable value of the setting has mechanisms that introduce an overhead which introduces the question: How significant is this overhead compared to best effort and how does it impact the latency especially and what implementations perform best when using each value?

The first paper, \cite{dds_a_performance_comparison_of_opensplice_and_rti_implementations_2013}, ran two types of tests that although had different reliability values, also had different values for other QoS settings. The first test experimented with one million packets and best effort whilst the second test looked at 5,000 packets and a reliable reliability. The difference in the number of packets significantly affects the results and does not explicitly show us the impact of the reliability quality of service since the authors intended to measure the throughput and samples per second for the first type of test whilst focusing on the latency for the second type of test. The results state that the round-trip time of the test where 5,000 packets were sent using the reliable value was lower than the test where one million packets were sent using best effort. This is to be expected but exactly how different was the round-trip time? We cannot answer this question as the authors of the paper do not share the exact measurements or the round-trip times of the first test (one million packets using best effort) as they focused on the throughput metrics as mentioned. 

Another paper, \cite{exploring_the_performance_of_ros2_2016}, did a better job at identifying the impact of best effort versus reliable although multiple QoS values were also set during the experiments (confounding the interpretation of the results on how best effort affects the performance of DDS). The paper experimented on RTI DDS and OpenSplice in a remote as well as a local environment. The results shown in Table 4 of the mentioned paper explain that it is impossible for RTI DDS to use reliable data greater than 128 kilobytes whilst OpenSplice can. The authors mention that this is due to a deficiency of additional configurations for large data on RTI DDS. These results were obtained when running the DDS implementations remotely. Furthermore, when ran remotely with best effort the results for both RTI DDS and OpenSplice were the same. Data transport was possible for all data lengths except for 2 and 4 megabytes where the data transport was still possible but was missing the deadline.

When running the tests locally (where messages were passed over a local loopback), it was again impossible for RTI DDS to use data greater than 128 kilobytes whilst OpenSplice was able to achieve this with the same reason mentioned before. The rest of the results were the same as mentioned before except when 4 megabytes of data being sent on RTI DDS using best effort was impossible due to a halt of process or too much data loss. Figure 7 of the mentioned paper compares the latencies of OpenSplice reliable and RTI DDS best effort and the results are quite intriguing as the graph shows us that the latency of RTI DDS best effort is in fact larger than when using OpenSplice reliable meaning that the best effort quality of service took longer. The difference increases as the data lengths increase though we cannot exactly deduce that best effort produces a larger latency as different DDS implementations were used and we cannot cleanly compare the results as they tend to implement the reliability mechanisms in different ways. This trend of RTI DDS best effort having a higher latency than OpenSplice reliable is shown throughout the paper in figures 8, 9 and 10 as well. With these interesting results we must investigate further how the reliability QoS affects the performance of DDS as well as compare between multiple implementations since they may implement the mechanisms differently in future work.

\textit{4.2.2.3. Number of Subscribers}\\
\label{textit:4.2.2.3.number_of_subscribers}
Overall, 2 papers evaluated the performance of DDS when varying the number of subscribers ranging from 1 subscriber to 45. The number of subscribers that are present can affect the performance of DDS to a varying degree depending on other QoS parameters. On top of this, we intend to find out what DDS implementations excel in terms of scalability. It would be useful to first identify the impact of the number of subscribers with the default QoS settings and then to experiment with various QoS values alongside changing the subscriber amount. However, in terms of what has been done already, \cite{data_distribution_services_performance_evaluation_framework_2018} experiments with 1, 9 and 45 subscribers in 3 separate tests. In the first test with 1 publisher and 1 subscriber it is seen in Figure 4b of the mentioned paper that OpenSplice takes the longest to complete the experiments throughout. Meanwhile, it is quite clear that FastRTPS has the second longest duration followed by OpenDDS with the fastest time up to 10 kilobytes. When compared with the second test where 9 publishers and subscribers are present, OpenSplice still takes the longest to complete the tests throughout all data lengths while OpenDDS and FastRTPS seem to have similar measurements in terms of how long the tests took up to 10 kilobytes. Between 10 kilobytes and 1 megabyte, FastRTPS took longer than OpenDDS to send the packets. In the final test where 45 publishers and subscribers are present, it is again shown that OpenSplice take the longest. However, throughout the different data lengths, both FastRTPS and OpenDDS have similar durations and from these results we can see that OpenSplice does not handle scalability well and the authors of the mentioned paper also state that OpenSplice has no ability to run a lot of participants as it has a limit of 120 participants. From this we can determine that OpenSplice is not designed for larger systems with numerous participants and large data lengths.

Scalability focused tests were ran in \cite{performance_assessment_of_omg_compliant_data_distribution_middleware_2008} comparing RTI DDS and OpenDDS where they experimented with 1, 2, 4, and 7 subscribers. For 1 and 2 subscribers, both RTI DDS and OpenDDS had almost the exact same one-way transit time whilst for 4 and 7 subscribers, RTI DDS outperforms OpenDDS in terms of latency. From these results we can determine that RTI DDS handles scalability better than OpenDDS for 4 or more subscribers up to 7 subscribers.

\hfill
\subsubsection*{4.2.3. Evaluation of a Single DDS Implementation}
\label{subsubsection:4.2.3.evaluation_of_a_single_dds_implementation}
\hfill\\
The papers mentioned within this section have evaluated a single DDS implementation to either evaluate the performance of DDS within a certain scenario, compare novel solutions with traditional DDS solutions in terms of performance or even applying DDS to alternative standards and then evaluating its performance. The usefulness of this specific segment of the literature reviewal may result in a reinforcement of what has been previously stated in section \nameref{subsubsection:4.2.2.evaluation_of_multiple_dds_implementations} or the results attained may even be significantly different and will therefore produce new questions.

\textit{4.2.3.1. Evaluation of DDS using Default QoS Settings}\\
\label{textit:4.2.3.1.evaluation_of_dds_using_default_qos_settings}
DDS has many competitors in terms of network communication standards and evaluating the performance of a system using the default QoS settings as stated by the OMG standard provides an adequate foundation for identifying the base performance of DDS. One can then tune these parameters to identify performance benefits or detriments and we intend to discover what has been identified in terms of the base performance of DDS using the default settings. However, before we undertake this discovery we first need to identify what exactly these default settings are and have therefore listed the exact parameter values in Table \ref{default_qos_value_table}.

\begin{table}[h]
    \begin{center}
        \begin{tabular}{|p{3cm}|p{5cm}|}
            \hline
            \textbf{Parameter} & \textbf{Default Value}\\
            \hline
            Durability & volatile \\
            Deadline & infinite \\
            Latency Budget & 0 \\
            Reliability & reliable \\
            History & keep\_last \\
            \hline
        \end{tabular}
    \end{center}
    \caption{Default QoS parameter values.}
    \label{default_qos_value_table}
\end{table}

There are other parameters where the mentioned papers have evaluated multiple values such as data length, total messages sent, total duration of the test, number of publishers and subscribers, and the locations where these publishers and subscribers reside, i.e., how many machines exist and what participants they accommodate for. With this in mind, we can analyse the performance of DDS according to various data lengths before identifying how the performance is affected over time throughout the duration of the test.

When a subscriber and a publisher are ran on the same local machine with the default QoS settings and 500 byte messages are disseminated, \cite{evaluating_a_prototype_approach_to_validating_a_dds_based_system_architecture_for_automated_manufacturing_environments_2014} shows us in Figure 6 of the paper that the highest latency attained is 50 milliseconds with an average latency of around 25 milliseconds for the OpenDDS implementation. Meanwhile, \cite{qos_aware_real_time_pub_sub_middleware_for_drilling_data_management_in_petroleum_industry} looked at implementing DDS within the petroleum industry and ran tests using the default QoS settings as mentioned using the RTI DDS implementation and the results when compared to the previous paper are similar. For 512-byte messages, the results show an average latency of 18 milliseconds which is lower than that of OpenDDS (25 milliseconds).

When varying the data lengths, figure 8 of the mentioned paper shows that the latency increases exponentially after being constant up to 256 bytes and this trend was present when experimenting with one publisher and multiple subscribers. This pattern was also seen in Data Distribution Service for Industrial Automation where the latency was constant up to 512 bytes before exponentially increasing.

To summarise, on average, latencies seem to be constant up to 256 bytes before exponentially increasing while for message sizes of 512 bytes, an average latency between 18 to 25 milliseconds is achieved.

\textit{4.2.3.2. Performance Impact of Reliability}\\
\label{textit:4.2.3.2.performance_impact_of_reliability}
One of the most popular experiments carried out throughout the analysed literature includes the comparison of performance when the reliability is set to “best effort” versus “reliable”. Theoretically, the “best effort” reliability should provide a lower latency as there are no mechanisms that introduce overheads confirming message dissemination. However, in this section we aim to investigate how significant the influence of the introduced performance overhead is and therefore will become a vital deciding factor that may lead to the usage of a reliable setting.

Figure 6 of \cite{an_investigation_on_the_applicability_of_dds_middleware_as_a_systems_integration_tool_2011} shows us that reliable and best effort values of the same QoS setting combinations seems to have similar results in terms of latency when ran on the RTI DDS implementation.

Figure 10 of \cite{data_distribution_service_for_industrial_automation_2012} indicates that for data lengths between 16 bytes to 128 bytes, reliable has a higher latency as expected. For messages of 8 bytes, it seems that the latency produced from both best effort and reliable seem to be almost the same and this also seems to be the case for 256-byte messages as well as 4096-byte messages. In terms of throughput, \cite{qos_aware_real_time_pub_sub_middleware_for_drilling_data_management_in_petroleum_industry} shows us in Figure 11 that for 2 and 4 subscribers, both reliable and best effort produce the same throughput. For any more subscribers, the reliable reliability produces more throughput as expected.

To summarise, theoretically we expect the latency and throughput of reliable tests to be higher than that of best effort. This has proven to be true in most cases, though for messages of 8, 256, and 4096 the latency seems to be almost the same for both reliable and best effort. Furthermore, when up to 4 subscribers are present, the effects of the reliable value is not shown.

\textit{4.2.3.3. Performance Impact of Quality of Service Tuning}\\
\label{textit:4.2.3.3.performance_impact_of_quality_of_service_tuning}
Two papers have looked at tuning the QoS settings to suit their scenarios to gain the best performance. We can therefore collect the results from these results to determine how the changes in QoS affected the performance. 

The first paper, \cite{al_yad_a_wearable_sensor_network_over_dds_middleware_for_industrial_application_2015} ran tests with a durability of "volatile", reliability of "reliable", liveliness of "automatic", history of "keep\_last" with a depth of 5 (instead of 1), owner of "exclusive" (instead of shared), owner strength of 5 (instead of 0), and a minimum separation value of 0.2 seconds (instead of 0) whilst using RTI DDS. A total of 100,000 messages of 28 bytes in length were sent and the performance both with and without the QoS settings were compared. In terms of throughput, the results were very similar, and the differences were not noticeable. The tuned QoS tests had an average throughput of 769 Mbps whilst the default settings produced an average throughput of 778 Mbps. A slightly better throughput. When looking at the latency results, of the 10 iterations, 5 produced lower latencies with the tuned QoS settings whilst the other 5 produced higher latencies. It is worth noting that on the 6th and 7th iteration, the tuned QoS tests produced the highest latencies of 5500 to 7000 microseconds whilst the default QoS parameters had latencies of 3500 and 3200 microseconds respectively. Figure 7 of the paper shows that these experimental results seem to be outliers.

Another paper, \cite{dds_based_implementation_of_smart_grid_devices_using_ansi_c12.19_standard_2017}, evaluated the performance using a durability of "transient\_local", a reliability of “best effort” on the publisher and “reliable” on the subscriber, a history of "keep\_all" on the publisher and "keep\_last" on the subscriber, and unlimited resource limit on the publisher and a value of 1 for the subscriber. The tests also compared between local area network (LAN) and wide area network (WAN) and increased the number of publishers and subscribers up to 6. The results for the default QoS settings show a lower latency throughout all publisher and subscriber amounts. There exists a difference of around 5 milliseconds between the default and non-default QoS settings with the default QoS producing lower latencies which is an outrageous difference. Overall, the latency linearly increases as the participants increase and these results are the same for the WAN tests. The throughput follows the same pattern though the default QoS settings has less throughput in both the LAN and WAN cases.

\textit{4.2.3.4. Performance Impact of Scalability}\\
\label{textit:4.2.3.4.performance_imapct_of_scalability}
These reviewed papers have investigated the effect on the performance of DDS in terms of scalability which includes varying the number of publishers, subscribers, DataWriters, and DataReaders.

Throughout the reviewed literature, two papers have investigated the effects of adding participants. The first, \cite{benchmarking_publish_subscribe_middleware_for_radar_applications_2007} experimented with values of data lengths ranging from 128 to 32768 bytes and increased the number of subscribers up to 7 whilst using a true multicast QoS setting. The results show that the median latencies increase linearly as the number of subscribers increase for data lengths up to 8192. Furthermore, the increase for 32768-byte messages in terms of latency from 1 subscriber to 4 was greater than from 4 subscribers to 7 and it must also be noted that the 32768-byte messages had outrageous maximum latencies of around 5000 microseconds. Moreover, when 2 subscribers were used, the 128-byte messages had a higher maximum latency that all other lengths up to 8192 bytes. As well as this, 4 kilobyte messages with 7 subscribers had a maximum latency of 2000 microseconds while 8 kilobyte messages had a maximum latency of just over 1000 microseconds.

The second paper, \cite{performance_evaluation_of_dds_based_middleware_over_wireless_channel_for_reconfigurable_manufacturing_systems_2015} ran tests using messages of 1024 bytes, 1 publisher and various subscriber amounts. For the experiments where 1 publisher is used, the latency increases linearly with the number of subscribers. When 2 to 4 publishers were active the latency did not follow the same pattern as before. In fact, the latency decreased when a second subscriber was added and then continued to increase for all other additional subscribers. In terms of the throughput, 150-byte messages were transported and in the case of 1 publisher and many subscribers, the throughput seemed to remain constant throughout the increase of subscribers with a slight decrease from 6 subscribers to 8. Figure 9 of the mentioned paper shows that when multiple publishers are used the increase in publishers seems to increase the throughput when comparing 2 publishers to 4. It seems an increase in subscribers does not affect the throughput.

\textit{4.2.3.5. Other Experimental Findings}\\
\label{textit:4.2.3.5.other_experimental_findings}
Other experimental findings are those that have not exactly been repeated in multiple papers. These tend to be papers that have carefully tuned the QoS parameters to suit specific scenarios that the authors were interested in.

Evaluation of 1 publisher with 7 subscribers was done in \cite{a_study_of_publish_subscribe_middleware_under_different_iot_traffic_conditions} and looked at the effects of the multicast QoS setting. The paper mentions that the results show an improvement in application performance ranging from 272\% to 842\% resulting from several experiments. The same paper also investigates the impact of TurboMode and AutoThrottle; two RTI implementation specific features. According to \cite{batch_qospolicy}, "Turbo Mode is an experimental feature that uses an intelligent algorithm that adjusts the number of bytes in a batch at runtime according to current system conditions, such as write speed (or write frequency) and sample size. This intelligence is what gives it the ability to increase throughput at high message rates and avoid negatively impacting message latency at low message rates". With AutoThrottling, the publisher automatically changes the writing rate based on the number of unacknowledged samples in the send queue to avoid blocking. The results of the experiments of the paper found that in 1-to-1 tests using TurboMode, the throughput improved 24 to 850\% and in 1-to-7 tests, the throughput improved 8\% to 1167\%. With the AutoThrottle tests, the 1-to-1 results show that the throughput was reduced by 18\% and the latency was also reduced by 3.6\% whilst with the 1-to-7 tests, the throughput was reduced by 19\% and the latency by 8.8\%.

Performance of DDS when both on LAN and WAN was carried out in \cite{an_investigation_on_the_applicability_of_dds_middleware_as_a_systems_integration_tool_2011}. Whilst alternating reliability settings where the results have been mentioned in section 4.2.3.2, the paper alternates between using a 100 Mbps LAN ethernet versus a 54 Mbps WAN connection. The results show that for messages of 28 bytes in length, a minimum latency of 375 microseconds is attained on the LAN connection. Furthermore, the paper mentions that the WAN experimental results tend to have high variance whilst the LAN experimental results were much more stable. From these results we can conclude that LAN tests are preferred due to its stability though WAN is capable of being used when required.

Evaluation of the performance of DDS when varying values of the latency budget was carried out in \cite{data_distribution_service_for_industrial_automation_2012}. The authors experimented with values between 0 and 100 and the results of the experiments found that a higher latency budget led to a higher average latency with only a difference of 40 microseconds. The case was the opposite for the maximum latency values. It was also found that a higher latency budget led to higher jitter.

Overall, we have seen from the non-categorised experiments on individual DDS implementations that using multicast improves application performance with multiple participants, WAN measurements have high variance compared to LAN, a higher latency budget leads to a higher average latency, lower maximum latency, and a higher jitter, TurboMode improves throughput, and finally, AutoThrottle reduces the latency at the cost of throughput.

\hfill
\subsubsection*{4.2.4. Evaluation of Unnamed DDS Implementations}
\label{subsubsection:4.2.4.evaluation_of_unnamed_dds_implementations}
\hfill\\
A limited amount of the literature evaluated the performance of DDS though the authors had not specifically stated the implementation that they used or may have used simulations. These papers have been included in this section and a total of 4 papers were identified under this category.

\textit{4.2.4.1. Evaluation of Scalability on DDS using Simulations}\\
\label{textit:4.2.4.1.evaluation_of_scalability_on_dds_using_simulations}
Two papers investigated the effects of DDS under network simulations whilst increasing the number of participants for each test. It should be stated that simulation performance may not be the same as the performance from a physical setup and should therefore not be compared. The piece of literature \cite{router_design_for_dds_architecture_and_performance_evaluation_2015} ran tests with 1024-byte messages whilst increasing the number of subscribers from 20 to 100. The results show that the latency increases linearly with the number of subscribers with the increase between 20 and 60 subscribers being greater than the increase between 60 and 100 subscribers. The results also show that the latency ranges from 5 milliseconds at 20 subscribers to 7 milliseconds at 100 subscribers. There is an almost linear increase in the number of messages in correspondence with the increase in subscribers. At 20 subscribers there are around 1000 messages being sent whilst at 100 subscribers there are around 13,500 messages being sent which is a much larger increase than what was seen in the latency.

Another paper, \cite{a_performance_simulator_for_dds_networks_2015} evaluated the performance of DDS when varying the number of DataWriters and DataReaders over 50 and 100 participants. 8 scenarios were run increasing the number of DataWriters and DataReaders by 50 per scenario resulting in the first scenario having the least number of DataWriters and DataReaders and the eighth scenario having the greatest number of DataWriters and DataReaders. The results show a strange pattern. Scenario 5, with 100 participants, 200 DataWriters and 200 DataReaders had the lowest throughput in terms of messages per participant whilst Scenario 4 with 50 participants, 350 DataWriters and 350 DataReaders had the highest throughput. A linear increase is then seen from scenario 5 up to scenario 8. It seems that the throughput shown in terms of messages per participant increases with the amount of DataWriters and DataReaders.

In summary, 2 papers have evaluated the performance of DDS whilst increasing the number of participants or DataWriters/DataReaders. Generally, the latency increases with the number of subscribers and so does the throughput though much more significantly. When adding DataWriters and DataReaders  we see a strange pattern where 100 participants with 200 DataWriters and DataReaders produces a lower throughput than 50 participants with 350 DataWriters and DataReaders though the throughput generally increases with the number of DataWriters/DataReaders.

\textit{4.2.4.2. Evaluation of the Impact of Multicast}\\
\label{textit:4.2.4.2.evaluation_of_the_impact_of_multicast}
Two papers also investigated the impact of the use of the multicast QoS setting. Both papers ran tests with the exact same configurations (though one varied the messaging rate from 50 to 100 samples per second). Both papers sent a total of 10,000 messages using the best effort reliability and varied the message sizes from 32 to 63,000 bytes.

Focusing on the unicast results, \cite{dds_based_interoperability_framework_for_smart_grid_testbed_infrastructure_2015} showed a very shallow linear increase up to 8192 bytes before an exponential rise from 8192 to 63,000 bytes in terms of latency. The second paper, \cite{a_dds_based_energy_management_framework_for_small_microgrid_operation_and_control_2018} also shows a linear increase in the latency from 32 to 8192 bytes with an exponential rise from 8192 to 63,000 bytes.

The first paper mentioned found similar patterns as the unicast results within its multicast results with a higher latency at 63,000 bytes. The second paper shows that the use of multicast provides a constant latency up to 512 bytes with a linear increase from 512 to 8192 bytes followed by an exponential rise from 8192 to 63,000 bytes which is very similar to the unicast results. In terms of the actual latency values being impacted by multicast, the results show that using multicast seems to increase the latency very slightly overall for all data lengths in both papers with the second paper showing a larger increase.

\hfill
\subsubsection*{4.2.5. Evaluation of DDS Implementations in a Virtualised Environment}
\label{subsubsection:4.2.5.evaluation_of_dds_implementations_in_a_virtualised_environment}
\hfill\\
This section will look at papers that have evaluated the performance of DDS whilst using virtual machines to deploy the DDS implementations. A total of 5 papers have been categorised under this section though 3 of these papers were written by Marisol Garcia-Valls, Rosbel Serrano-Torres, and Pablo Basanta-Val and include the same content. These three papers are \cite{analyzing_point_to_point_dds_communication_over_desktop_virtualization_software_2017}, \cite{benchmarking_communication_middleware_for_cloud_computing_virtualizers} and \cite{performance_evaluation_of_virtualized_dds_middleware}.

These papers ran tests with default quality of service parameters, 1 publisher, and 1 subscriber and the focus of these papers was to identify the effects of virtualisation on DDS. The results show that in terms of invocations per second, RTI has the least number of invocations per second for all the types of tests and this is less than OpenSplice's. Generally, small data had more invocations per second than medium data both virtualised and non-virtualised. In terms of virtualisation overhead, the results dictate that small data has more virtualisation overhead as well as the fact that RTI has more virtualisation overhead than OpenSplice for both small and medium data. When looking at the results in terms of blocking time, its seen that RTI has a greater blocking time in general than OpenSplice though virtualisation significantly increases the blocking time for both implementations.

Overall, it seems that OpenSplice is better suited for virtualisation than RTI DDS from these results since RTI has less invocations per second, more virtualisation overhead, and a greater blocking time.

Tests were run on virtual machines in \cite{cloud_iec_61850_dds_performance_in_virtualized_environment_with_opendds_2017}. A total of 11,000 messages of 250 bytes were sent during the test along with multicast. The results in the graphs of the paper show an average latency in the range between 270 and 320 microseconds. The paper states that the maximum latencies vary from 1 to 5 milliseconds but are less than one percent of the total samples and the quantile graphic shows that more than 95\% of the samples are very close to the median.

\section{Discussion}
\label{section:5.discussion}
\subsection*{5.1. General Discussion}
\label{subsection:5.1.general_discussion}
This paper has analysed the literature and identified the results for several categories; comparison of DDS with other publish-subscribe technologies, evaluation of single, multiple, and unnamed DDS implementations, and evaluation of DDS within a virtualised environment. In this section we intend to generalise on the findings from the previous section and to identify any points of interest. It is vital to state that any products that are described as having a better performance is only a result of the survey of the mentioned literature throughout the paper and does not explicitly mean that the certain product will always behave in the described way.

When comparing DDS with other publish-subscribe technologies we first see that the performance of DDS is outperformed by traditional socket-based communications though the minimum latency observed for DDS states otherwise. Compared with HLA, DDS generally outperforms in all regards, as DDS has smaller jitter, higher throughput, and slightly lower latencies. This is also the case with MQTT. DDS consistently outperforms MQTT in terms of throughput and the latency of DDS outperforms MQTT for 64-byte messages though ZeroMQ's latency does outperforms DDS's. In summary, it seems that out of the evaluated technologies, only traditional socket-based communication and ZeroMQ outperform DDS whilst DDS generally outperforms HLA, MQTT, RabbitMQ, and AMQP according to the surveyed literature and the enclosed experiments.

As mentioned before, to compare multiple DDS implementations, its best to narrow the results down according to each configuration parameter. Starting with the data length, the results show us that OpenSplice and OpenDDS are better suited for small data whilst RTI DDS and FastRTPS are better suited for larger data. When looking at reliability, the literature has shown that OpenSplice can use reliable data over 128KB messages whilst RTI DDS cannot. On top of this, RTI DDS seems to lack the ability to send 4MB messages using best effort and it’s also shown that RTI DDS's latency when using best effort is greater than OpenSplice’s latency when using reliable though it’s not recommended to compare these results as they are different implementations with different reliability settings. When looking at scalability and increasing the number of subscribers, OpenSplice is not well-suited for scalability taking the longest to complete all tests. OpenDDS and FastRTPS generally have similar results and are faster than OpenSplice. RTI DDS outperforms OpenDDS in terms of latency for subscribers ranging from 4 to 7. Therefore, it seems that RTI DDS may be the most performant when focusing on scalability though an inadequate amount of experiments have been carried out to solidify this hypothesis as well as the fact that it is quite difficult to justify this increase in scalability as the gap is minimal jumping from 4 to 7 subscribers compared to a significant increase in the tens or hundreds.

We can use the results from the papers that evaluated a single DDS implementation to solidify what has been presented for certain implementations and to also see how various quality of service settings affect the performance of DDS when their default values are utilised. For these default settings, it seems that RTI DDS has a lower average latency than OpenDDS. From these results we cannot confidently say that RTI DDS outperforms OpenDDS though from what we've seen, it does for default quality of service settings. Looking at the effects of reliability on the performance of DDS is quite hard as the effects are not clear. The literature has shown that reliable and best effort seem to produce similar results as well as the fact that the reliable value increases the throughput for more than 4 subscribers. Finally, when focusing on scalability, it’s strange that the increase in latency from one subscriber to four was greater than the increase in latency from four subscribers to seven. This could mean that as more subscribers are added, the latency increase is reduced. The results also show that when 7 subscribers are used, 4KB messages had a higher maximum latency than 8KB messages. Furthermore, the results indicate that increasing the number of subscribers does not affect the throughput which is opposite to our theoretical understanding since increasing the number of subscribers ensures that more data must be sent which should therefore lead to an increase in the throughput though we cannot state that this reason is concrete as this depends on how loaded the network is as well as other factors.

Again, we can find the results useful when evaluating unnamed DDS implementations to identify how DDS performs and when looking at the scalability of DDS under simulations its seen that the latency only increases by 2 milliseconds when the subscribers increase by 80 whilst the number of messages sent increases by 12,500. This indicates that the throughput drastically increases which is the opposite of what was stated/seen in the previous paragraph. Multicast does not seem to significantly impact the performance of DDS. Overall, adding multicast seems to increase the latency instead of decreasing it, though this statement is only proven by two papers and should therefore be taken lightly. Finally, in terms of the performance of DDS under virtualised environments, OpenSplice is better suited for being run in a virtualised environment than RTI DDS as OpenSplice in a virtualised environment has more invocations per second, less overhead, and less blocking time even though for both implementations, virtualisation takes up a significant portion of the total execution time.

\subsection*{5.2. Research Gaps}
\label{subsection:5.2.research_gaps}
Although there are numerous pieces of literature that evaluate the performance of DDS, whether by itself, compared to other publish-subscribe technologies, or comparisons of its implementations, there exist still  research gaps in terms of the performance evaluation tests that could be carried out.

Reliability is a QoS parameter that has been experimented with in multiple papers though there are gaps within the surveyed literature research. Only two papers evaluate the effects of reliability, one of which uses non-default quality of service parameter values which may cloud the exact performance impacts of the reliability parameter and the other compares alternate values of the reliability setting whilst using different number of packets sent during the test (1,000,000 best effort packets versus 5,000 reliable packets). Not one piece of literature has explicitly evaluated the effects of the alternate values of the reliability setting whilst keeping values of other configuration parameters constant.

Scalability was another test parameter that was evaluated in a couple of papers with research gaps identified. The impact of 1, 9, and 45 publishers/subscribers was evaluated in \cite{data_distribution_services_performance_evaluation_framework_2018} on OpenDDS, OpenSplice, and FastRTPS whilst \cite{performance_assessment_of_omg_compliant_data_distribution_middleware_2008} evaluated 1, 2, 4, and 7 subscribers using RTI DDS and OpenDDS. From these numbers we can identify that RTI DDS hasn't been evaluated with more than 7 subscribers and 45 participants is the maximum number of participants evaluated. Furthermore, scalability has been evaluated using simulations and cannot fairly be compared to experimental evaluations which is another gap in this research domain.

Moreover, only 2 papers look at the effects of the multicast quality of service parameter and no literature has investigated the performance effects of the Durability, Deadline, Latency Budget, and History quality of service parameters which all theoretically could affect the performance of DDS.

Finally, majority of the reviewed literature does not look at the distribution of measurements and are instead based on the point estimates, which indicates a significant gap within the research.

\section{Conclusion}
\label{section:6.conclusion}
DDS has been used for almost two decades and its applications are numerous ranging from air traffic control systems to autonomous vehicles. Over the years DDS has been improved and various features have been added to it despite it already being flexible due to the ability to control various quality of service parameters and hopefully DDS continues to improve in the future. Investigating how DDS performs under what conditions can help us understand where its weaknesses and strengths lie and what one can do to improve this specification further. However, the literature evaluating the performance of DDS has been quite minimal and this has been identified within this survey paper. As well as this, majority of the papers do not look at distribution of measurements and are instead based on the point estimates which indicates a significant gap in the research.

In this paper, we have carried out a systematic literature review on the evaluation of the performance of DDS. From the hundreds of papers on DDS we identified 46 that mentioned performance of DDS and may or may not have included performance metrics. Applying our filtering, we narrowed these papers down to 29 from which we analysed the content and generalised on the themes and findings within the fourth and fifth section of this paper. The categories explored include comparison of DDS with other publish-subscribe technologies, evaluation of multiple DDS implementations, evaluation of a single DDS implementation, evaluation of unnamed DDS implementations, and evaluation of DDS implementations in a virtualised environment. We have extracted the various experimental performance tests carried out on DDS as well as the results from these experiments and have identified certain themes within the research as well as specific gaps discussed further in \nameref{subsection:5.2.research_gaps}.

This paper has only reviewed the performance of DDS under experimental evaluations and we believe that there is still more to be done in terms of evaluating the performance of DDS, such as model-based evaluations of DDS, experimental evaluations of security-enabled DDS, and model-based evaluations of security-enabled DDS. We believe that the content of this paper opens a path for further research into this field and that this research can help develop DDS further as well as aid users and practitioners in identifying whether DDS is the most performant communication technology to use, and more specifically which DDS vendor will be best suited for their use cases. In our future work, we plan to evaluate the performance of both security-enabled and vanilla DDS using both experimental and model-based evaluations.

\clearpage
\newpage

\section{Glossary}
\label{section:7.Glossary}


\printbibliography

\clearpage
\newpage

\section{Appendix}
\label{section:8.Appendix}

\begin{center}
    \begin{supertabular}[t]{|p{2cm}|p{6cm}|}
        \hline
        \textbf{Parameter} & \textbf{Description} \\
        \hline
        Deadline & Amount of time either between writing each sample on a DataWriter or reading each sample on a DataReader.\\
        \hline
        Destination Order & Controls what sample is chosen when multiple DataWriters send the same sample to a DataReader.\\
        \hline
        Durability & Controls whether data reader that join late can access already published data. \\
        \hline
        Durability\_Service & Only used when Durability QoS has value of \texttt{transient} or \texttt{persistent}.\\
        & Has multiple settings. \\
        & - \texttt{service\_cleanup\_delay} \\
        & - \texttt{history\_kind} \\
        & - \texttt{history\_depth} \\
        & - \texttt{max\_samples} \\
        & - \texttt{max\_instances} \\
        & - \texttt{max\_samples\_per\_instance} \\
        & \textbf{service\_cleanup\_delay} \\
        & How long to keep all information regarding an instance. \\
        & If set to 0 then the information is deleted immediately after being sent. \\
        & If set to INFINITE then information is kept forever. \\
        & \textbf{history\_kind} \\
        & Can either be \texttt{keep\_last\_history} or \texttt{keep\_all\_history}. \\
        & \texttt{keep\_last\_history}: Keep the last depth number of samples per instance. \\
        & \texttt{keep\_all\_history}: Keep all samples. \\
        & \textbf{history\_depth} \\
        & \textit{\texttt{history\_kind} must be set to \texttt{keep\_last\_history}.} \\
        & This is the amount of samples to keep per instance. \\
        & \textbf{max\_samples} \\
        & Max number of live samples that can be stored for a DataWriter or DataReader. \\
        & \textbf{max\_instances} \\
        & Max number of instances that can be managed by a DataWriter or DataReader. \\
        & \textbf{max\_samples\_per\_instance} \\
        & Max number of samples of any one instance that can be stored for a DataWriter or DataReader.\\
        \hline
        Entity\_Factory & Controls whether entities can be enabled or not for communication when created. \\
        \hline
        Group\_Data & Allows the application to attach additional information to the created Publisher or Subscriber. \\
        \hline
        History & Controls how many samples to be stored on either the DataWriter or DataReader. \\
        \hline
        Latency Budget & Provides a hint to the maximum acceptable delay from the time the data is written to the time it is received by the subscribing applications. \\
        \hline
        Life Span & Controls how long a sample can be stored in cache for being deleted. \\
        \hline
        Liveliness & Controls whether a DataWriter is "alive" in order to be used by the ownership parameter. \\
        \hline
        Ownership & Controls whether the service allows multiple DataWriters to update the same instance. Can take the values of \texttt{shared} or \texttt{exclusive}. \\
        & \texttt{shared}: Multiple DataWriters can update the same instance. The subscriber will access all modifications of the instance. \\
        & \texttt{exclusive}: Each instance can only be modified by only one DataWriter. If there are multiple DataWriters then the strongest one is chosen to be the only one that can modify the instances. Strength is determine by the value of the \texttt{ownership\_strength} QoS. \\
        \hline
        Partition & A method of further categorising participants within a domain. \\
        \hline
        Presentation & Controls how the data arrives to the DataReader including the sequence of the samples. \\
        & There are 3 settings: \\
        & - \texttt{access\_scope}\\
        & - \texttt{coherent\_access} \\
        & - \texttt{ordered\_access} \\
        & \textbf{access\_scope} \\
        & \textit{Only works if \texttt{coherent\_access} is set to true.}\\
        & Controls how the samples are ordered (by topic, instance, or both). \\
        & Can be of three types: instance, topic, or group. \\
        & \textbf{Instance}: Queue is ordered/sorted for every instance. \\
        & \textbf{Topic}: Queue is ordered/sorted per topic. \\
        & \textbf{Group}: Queue is ordered/sorted per topic across all instances belonging to DataWriter/DataReader within the same publisher/subscriber. \\
        & \textbf{coherent\_access} \\
        & Controls whether the service will preserve the groupings of changes made by the publishing application by means of the operations \texttt{begin\_coherent\_change} and \texttt{end\_coherent\_change}. \\
        & \textbf{ordered\_access} \\
        & Controls whether the service will preserve the order of the changes. \\
        \hline
        {\scriptsize Reader\_Data\_Lifecycle} & Controls what happens to the data in the receive queue of the DataReaders. There are two main aspects: \\
        & - when DataWriters no longer exist \\
        & - automatic disposal of samples \\
        & \textbf{When DataWriters No Longer Exist} \\
        & If the DataWriter has sent a sample that no longer exists, the DataReader will wait for {\small \texttt{autopurge\_nowriter\_samples\_delay}} duration and then disposes of the samples sent by the now non-existent DataWriter. \\
        & \textbf{Automatic Diposal of Samples} \\
        & The DataReader can automatically dispose of samples within it's receive queu after waiting for {\small \texttt{autopurge\_disposed\_samples\_delay}} duration. If the user would like the samples to be kept on the receive queue then this duration should be set to infinite (which is the default value). \\
        Reliability & Controls whether missed samples are resent or not. There are two types of reliability: \\
        & - \texttt{best\_effort}\\
        & - \texttt{reliable}\\
        & \textbf{best\_effort}\\
        & Any missed samples are not resent. The order of samples in the history of the DataWriter will match the history of received samples on the DataReader. \\
        & \textbf{reliable} \\
        & Any missed samples are resent. If the DataReader does not receive a sample then the DataWriter blocks the future sent samples to solve the issue. This means that the DataWriter will block all future samples while it tries to resend the missed sample. This blockage time can be controlled via the \texttt{max\_blocking\_time} setting. \\
        \hline
        Resource Limits & Defines what the maximum number of samples, instances, and samples per instance can be during the communication. By default it is set to unlimited. \\
        & There are three types of limits: \\
        & \texttt{max\_samples}: Sets the size and causes memory to be allocated for the send or receive queues. Default is unlimited.\\
        & \texttt{max\_instances}: The maximum number of instances that can be stored during communication. Default is unlimited. \\
        & \texttt{max\_samples\_per\_instance}: Used when topics are keyed. This represents the maximum number of samples with the same key that are allowed to be stored by a DataWriter or DataReader. Default is unlimited. \\
        \hline
        {\small Time\_Based\_Filter} & On the DataReader side, the DataReader specifies a time duration for which a single instace/sample must arrive. If more than one arrives then only one is taken and the rest are dropped. This time duration must be less than the \texttt{deadline} QoS. \\
        \hline
        Topic\_Data & Allows the application to attach additional information to the topic. \\
        \hline
        Transport Priority & Optional QoS parameter that describes the priority of the sample being sent.\\
        \hline
        User\_Data & Allows the application to attach information to entity objects so that the remote application can access this information. \\
        \hline
        {\scriptsize Writer\_Data\_Lifecycle} & Controls whether unregistered instances should be deleted or not. Instances can be unregistered via the \texttt{unregister()} operation.\\
        \hline
    \end{supertabular}
    \newline
    \captionof{table}{All 22 QoS settings defined in the DDS standard.}
    \label{all_qos_table}
\end{center}

\end{document}